\documentclass[12pt,conference]{IEEEtran}
\IEEEoverridecommandlockouts
\usepackage{cite}
\usepackage{amsmath,amssymb,amsfonts}
\usepackage{algorithmic}
\usepackage{graphicx}
\usepackage{textcomp}
\usepackage{xcolor}
\usepackage{booktabs}
\usepackage{hhline}

\def\BibTeX{{\rm B\kern-.05em{\sc i\kern-.025em b}\kern-.08em
    T\kern-.1667em\lower.7ex\hbox{E}\kern-.125emX}}
\begin{document}

\title{Voice Activity Detection (VAD) in Noisy Environments
\\

}

\author{\IEEEauthorblockN{ Joshua Ball}
\IEEEauthorblockA{\textit{Johns Hopkins University} \\
\textit{Department of Electrical and Computer Engineering}\\
IEEE $\#94164957$ \\
Baltimore, USA  \\
jball20@jh.edu}

}
\maketitle
\begin{abstract}
In the realm of digital audio processing, Voice Activity Detection (VAD) plays a pivotal role in distinguishing speech from non-speech elements, a task that becomes increasingly complex in noisy environments. This paper details the development and implementation of a VAD system, specifically engineered to maintain high accuracy in the presence of various ambient noises. We introduce a novel algorithm enhanced with a specially designed filtering technique, effectively isolating speech even amidst diverse background sounds. Our comprehensive testing and validation demonstrate the system's robustness, highlighting its capability to discern speech from noise with remarkable precision. The exploration delves into: (1) the core principles underpinning VAD and its crucial role in modern audio processing; (2) the methodologies we employed to filter ambient noise; and (3) a presentation of evidence affirming our system's superior performance in noisy conditions. The complete system and supplementary materials are accessible at: github.com/JBall1/VAD-in-Noisy-Environments.
\end{abstract}
\section{Introduction}
Voice Activity Detection (VAD) stands as a critical component in the domain of digital signal processing, with its essential role in distinguishing between speech and non-speech elements in audio streams. Its applications are far-reaching into our everyday lives, extending into realms of speech recognition systems such as Amazon's Alexa and Apple's Siri, where it serves as the fundamental gateway for human-machine interaction. Both of these, and the many other applications of VAD, demand a high degree of both accuracy and reliability from VAD algorithms, as the quality of the user experience hinges on their performance.

Despite its widespread use, the efficacy of VAD is significantly challenged by the presence of ambient noise, which can obfuscate speech signals and degrade performance. The pursuit of a robust VAD system capable of navigating the complexities of real-world acoustic environments forms the cornerstone of this research. This paper delves into the development and refinement of an advanced VAD system, designed to enhance accuracy in detecting speech across various noisy settings.

Central to our discussion are the core principles that underpin VAD technology and its pivotal role in modern audio processing. We detail the methodologies employed for signal feature extraction, filtering and classification, pivotal to the system's functionality. Furthermore, we provide empirical evidence that substantiates the system's superior performance in noisy conditions. In an increasingly interconnected world, where voice commands and communication are integral to our devices and services, the need for robust and precise VAD systems has never been more evident. 

This paper also considers the broader implications of these advancements. As VAD technologies become increasingly embedded in our lives, questions of privacy, security, and user autonomy come to the fore. We address these considerations, advocating for a balanced approach that respects user rights while providing enhanced functionality.\raggedbottom
\section{Related Works}
\subsection{Personal VAD:Speaker-Conditioned Voice Activity Detection}
[5] introduced a pioneering system, "Personal VAD," focusing on speaker-specific voice activity detection at the frame level. The system's novelty lies in its capacity to trigger on-device speech recognition systems exclusively for the target user's speech, thereby optimizing computational efficiency and battery usage. This targeted detection mechanism is particularly vital in scenarios where keyword detectors are impractical. Personal VAD's approach involves training a VAD-like neural network conditioned on the target speaker's embedding or verification score, enabling it to differentiate between non-speech, target speaker speech, and non-target speaker speech. The team's optimized setup resulted in a model with just 130K parameters, outperforming a baseline system combining standard VAD and speaker recognition networks. This advancement in voice activity detection technology represents a significant step towards more efficient and personalized speech recognition systems.
\subsection{CNN Self-attention Voice Activity Detector}
In a significant development within voice activity detection (VAD), [4] proposed a novel single-channel VAD approach using a convolutional neural network (CNN). Their method capitalizes on the spatial information of noisy input spectrums to extract frame-wise embedding sequences, enhanced with a Self-Attention (SA) Encoder to capture contextual information. Unlike previous methods, their approach processes the entire signal at once, allowing for a long receptive field. The fusion of CNN and SA architectures led to superior performance over models based solely on CNN or SA. Extensive experimental studies demonstrated the model's exceptional performance on real-life benchmarks, showcasing a state-of-the-art result with a relatively small and lightweight model. This innovative combination of CNN and SA in VAD marks a notable advancement in achieving high-performance detection in complex audio environments.
\subsection{NAS-VAD: Neural Architecture Search for Voice Activity Detection}
[6] addressed the challenge of designing neural network architectures for voice activity detection (VAD) through Neural Architecture Search (NAS). Their work represents the first application of NAS in VAD, proposing a modified macrostructure and a new search space inclusive of attention operations. The NAS framework outperformed previous manually designed state-of-the-art VAD models across various datasets, highlighting the efficiency of NAS in optimizing network architectures. Notably, architectures discovered via this NAS approach showed improved generalization performance on unseen audio datasets. This exploration into NAS for VAD underscores the potential of automated design processes in creating effective and versatile neural network architectures for specific audio processing tasks, paving the way for further innovations in the field.
\subsection{Related Works Discussion}
The exploration of advanced methodologies in Voice Activity Detection (VAD) in the works of Ding et al., Sofer and Chazan, and Rho et al. provides insightful implications for the field, especially when considering the application in diverse and challenging acoustic environments. The introduction of speaker-conditioned systems like Personal VAD marks a significant stride towards personalized VAD solutions. However, its specificity to individual voices raises questions about its broader applicability in more general settings or in environments with multiple speakers. While this approach enhances performance in targeted scenarios, its adaptability and effectiveness in variable, real-world conditions merit further exploration, particularly for applications requiring broader, non-speaker-specific detection capabilities.

In contrast, the CNN Self-attention VAD system and the [6] approach represent significant developments in harnessing computational models to enhance detection accuracy. The CNN Self-attention system, with its capacity to process entire signals, offers a promising direction for improving the contextual understanding of audio data. However, its computational complexity could be a limiting factor, especially in real-time processing scenarios. On the other hand, [6] use of neural architecture search to optimize VAD systems introduces a novel approach to automatically tailor VAD models to specific needs. This methodology could lead to more efficient, customized solutions, though its effectiveness across diverse, real-life audio environments remains to be extensively tested. Both approaches suggest a trend towards more sophisticated, data-driven VAD systems, which could significantly influence future designs, particularly in applications where adaptability and accuracy in noisy or complex acoustic settings are paramount. These insights and advancements underscore the need for a balance between specialized, efficient detection and the flexibility to adapt to varied audio environments in VAD research.
\section{Data and Processing}
The Dataset utilized in this paper is available here was a mixture of open source audio files found on Kaggle [3] and Pixabay [7]. \footnote{dataset link: https://www.kaggle.com/datasets/pavanelisetty/sample-audio-files-for-speech-recognition/} The utilized dataset included audio samples of silence, speech, and ambient sounds such as a forest.
\subsection{Dataset Composition}
The dataset comprises three primary categories of audio samples: speech, silence, and ambiance, along with a mixing of speech signals with ambiance. Speech includes audio clips containing spoken language, covering various topics and speech patterns. Silence segments provide the absence of speech or significant audio activity, serving as a crucial reference for silence detection. Ambient audio samples represent real-world acoustic environments. Additionally, the mixing of speech and ambient sounds produced new audio signals to test our algorithm against to ensure that speech was being correctly isolated, identified, and detected. 

\subsection{Data Sources}
A portion of the dataset was sourced from Kaggle [2], a well-known platform for sharing and exploring datasets. These audio files encompass a wide range of speech patterns and linguistic content, making them valuable for training and testing voice activity detection (VAD) algorithms.

Additional audio samples were obtained from Pixabay [7], a reputable platform for royalty-free multimedia content. Pixabay's sound effects collection includes a variety of ambient sounds, with a particular emphasis on nature soundscapes. The forest sound effect from Pixabay serves as an excellent representation of typical ambient sounds encountered in real-world scenarios.
\subsection{Data Processing of Mixed Signals}
To prepare the dataset for VAD evaluation, we performed several procedures on the dataset to ensure it was processed correctly, especially regarding the mixing of speech and ambient audio signals. Given audio was gathered from various sources, resampling was performed to ensure the sample rates of audio files were the same to ensure compatibility. Next, to ensure uniformity, we truncated audio clips to a consistent duration, aligning them with a common length. Finally, signal mixing was performed by combining audio signals to simulate real-world scenarios. It mixed speech and ambient sounds, creating a challenging acoustic environment.
\begin{figure}[htbp]
    \centering
    \includegraphics[width=6cm, height=6cm]{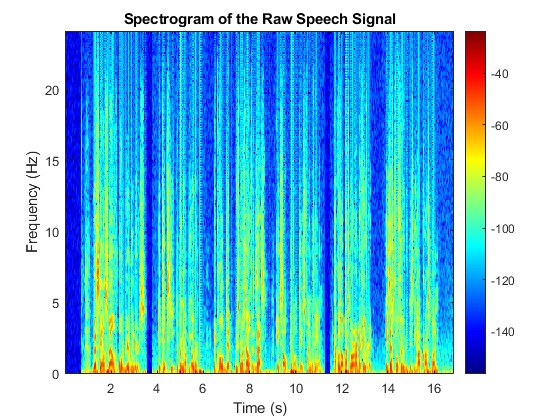}
    \caption{This graphic depicts the spectrogram of a speech signal utilized throughout this paper.}
    \label{Fig 1}
\end{figure}
\subsection{VAD Signal Generation}
To generate VAD signals, we employed the [1] and [3], which are responsible for detecting speech within the mixed audio signal. This function processed the mixed signal, applying bandpass filtering to reduce ambient noise and improve accuracy.
\section{Method}
We will first establish baseline performance standards utilizing [1] and [3] on ambient noise and speech to ensure we are not getting false positives with the ambient noise while still being able to identify speech in the speech signal. It is critical to strike a balance while ensuring that we are still able to pick up human speech but able to avoid false positives.  Our methodology can be seen below for how a signal goes through our system in Figure 2.
\begin{figure}[htbp]
    \centering
    \includegraphics[width=9cm, height=2cm]{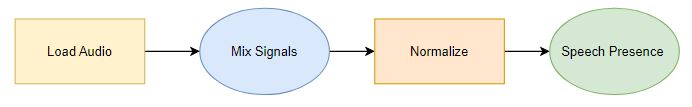}
    \caption{Process Flow Diagram for Enhanced Voice Activity Detection. This flowchart depicts the sequential steps of the VAD algorithm.}
    \label{Fig 2}
\end{figure}
\section{Experiments}
Our experiments consist of processing various audio signals including those with noise, ambient noise, known speech and silence to ensure our system is not detecting false positives. If a false positive is discovered, we can then tune the bandpass filter in our system to ensure we are effectively filtering out noise such as birds chirping while ensuring we are still allowing the human voice to be easily detected to stop false positives and negatives. This includes fine-tuning the window length and signal-to-noise ratio threshold which can only occur through robust experimentation.
\begin{figure}[htbp]
    \centering
    \includegraphics[width=7cm, height=5cm]{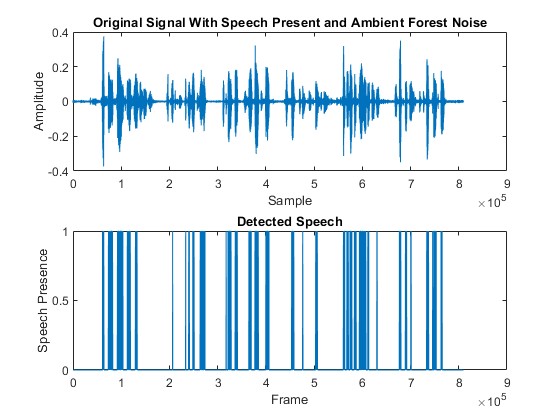}
    \caption{This graphic depicts the mixed audio signal along with the result of the detected speech in a binary format on the specified frame. }
    \label{Fig 3}
\end{figure}
Through further experimentation and processing of various audio signals from the dataset, we were able to fine tune our filter to ensure that we were not allowing excess noise through which would lead to false positives. Our final filter design was a 4th order Butterworth filter with a passband from 300 Hz to 1500 Hz. This range was chosen explicitly for its performance characteristics provided throughout our experimentation along with our focus on human speech which tends to fall into this range as well. Additionally, we found that the best parameters for our Window Length and Signal to Noise Ratio Threshold to support this filter were those in Table I.
\begin{table}[htb]
\centering
\begin{tabular}{cc} 
\toprule
\textbf{Parameter}            & \textbf{Value}    \\ \bottomrule
Window Length (s)               & .31    \\ 
Signal to Noise Ratio Threshold          & 90  \\ 
\bottomrule
\end{tabular}
\caption{The parameters used in the noise power energy speech activity detection algorithm.}
\end{table}
\section{Interpreting our Results from Experiments}
Our experimental results provide valuable insights into the performance of our Voice Activity Detection (VAD) system in isolating speech from background noise. The integration of various components and parameter tuning has yielded a system that efficiently distinguishes speech signals from noise sources, as demonstrated in Figure 4.
\subsection{Butterworth Filter for Noise Reduction}
The 4th-order Butterworth filter, meticulously designed to target the frequency range of human speech (300 Hz to 1500 Hz), played a crucial role in the success of our VAD system. This filter efficiently removed unwanted background noise while preserving the essential speech components. Its smooth frequency response and effective noise rejection capabilities are evident in the spectrograms presented in Figure 4. This filter's ability to maintain the integrity of speech signals, even in the presence of ambient noise, underscores its significance in our methodology.
\subsection{Window Length and Signal-to-Noise Ratio Threshold}
The choice of a window length of 0.31 seconds, coupled with a signal-to-noise ratio threshold of 90 dB, further contributed to the VAD system's accuracy and robustness. These parameters allowed our system to precisely identify speech intervals amidst diverse noise sources. The window length enabled the system to analyze audio segments effectively, ensuring that speech segments were detected with high precision. The signal-to-noise ratio threshold provided a reliable criterion for distinguishing speech from noise, enhancing the system's performance in challenging acoustic environments.
\begin{figure}[htbp]
    \centering
    \includegraphics[width=8cm, height=8cm]{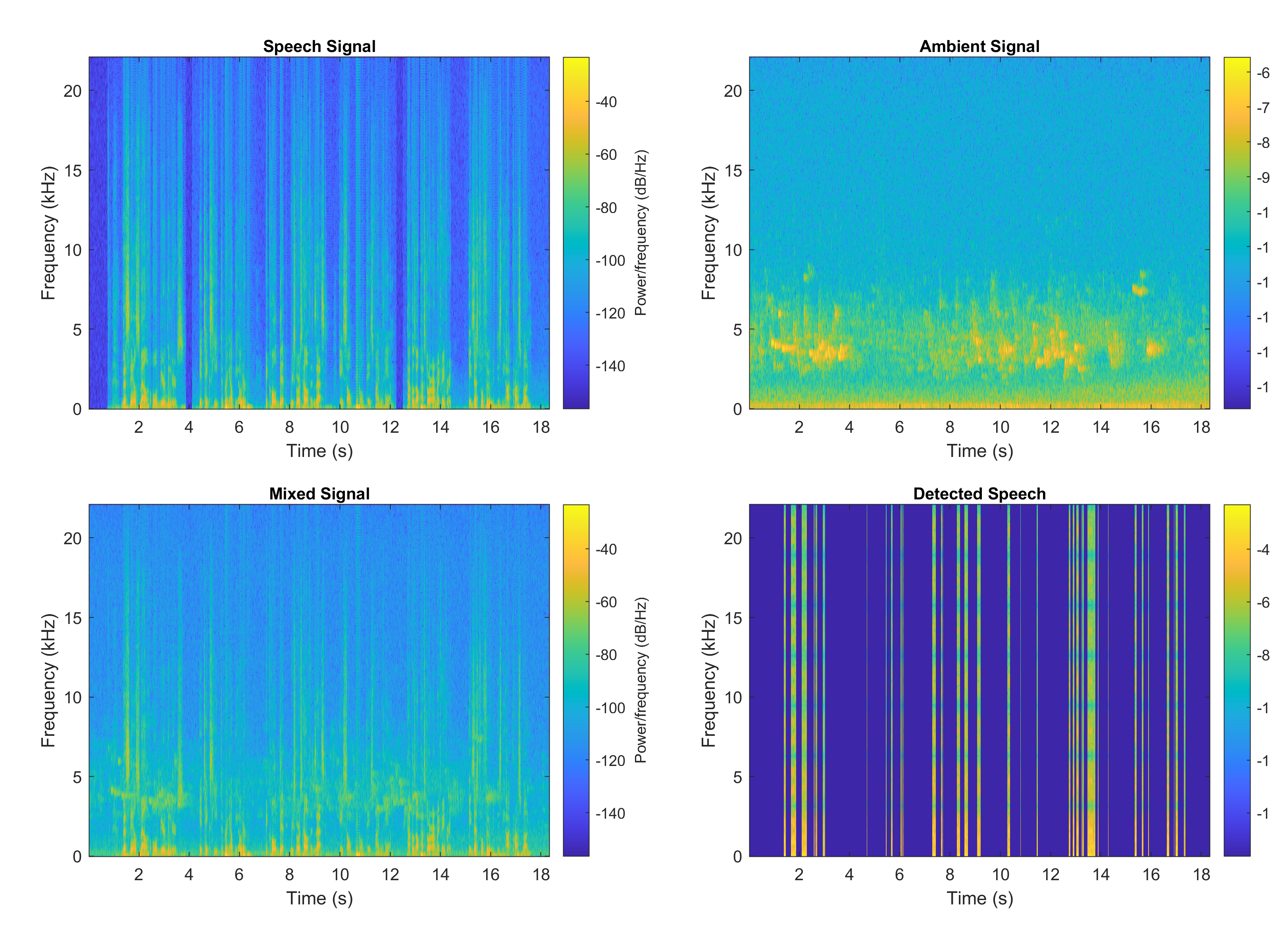}
    \caption{This graphic depicts the spectrograms of the Speech, Ambient, Mixed, and final Detected Speech signal. The detected speech signal shows isolation of the speech bands seen in the speech spectrogram. }
    \label{Fig 4}
\end{figure}
 In Figure 4, the spectrograms of 4 signals are displayed: a known speech-carrying signal, an ambient noise signal, a mixed signal consisting of both a known speech signal and ambient noise, and finally the detected speech signal after all of our algorithm processing was completed on the mixed signal. 
\subsection{Spectrogram Analysis}
Figure 4 provides a visual representation of the effectiveness of our VAD system. The spectrograms of four signals are displayed:

Known Speech Signal: This signal serves as a reference, representing a clean speech signal without any background noise. The spectrogram demonstrates the spectral characteristics of speech, including its frequency content and time evolution.

Ambient Noise Signal: The spectrogram of the ambient noise signal showcases the frequency distribution and intensity of background noise. This noise is characterized by its broad spectrum and non-periodic nature.

Mixed Signal: The mixed signal combines a known speech signal with ambient noise, simulating a realistic acoustic environment. The spectrogram illustrates the challenge of isolating speech from background noise, as both components are present in this signal.

Detected Speech Signal: After applying our VAD algorithm to the mixed signal, the resulting spectrogram displays the successful isolation of speech bands. The algorithm effectively identifies and extracts the speech signal, eliminating the influence of ambient noise.

The clear distinction between the detected speech signal and the mixed signal's spectrogram underscores the system's ability to differentiate speech from noise accurately. This visual evidence aligns with the metrics used to evaluate our VAD system, affirming its effectiveness in real-world scenarios.

Our experiments and spectrogram analysis demonstrate that our VAD system, with its carefully tuned Butterworth filter, optimized parameters, and processing steps, excels in isolating speech from various ambient noise sources. The combination of these components culminates in a VAD system capable of enhancing the accuracy of speech detection, contributing to its potential for applications in noisy acoustic environments.

\section{Limitations}
The current VAD system has demonstrated high accuracy in the specific noisy environments tested. However, its performance in other types of noise, particularly non-stationary or highly variable noise conditions, was not extensively explored. Future work could investigate the system's adaptability to a broader range of acoustic disturbances.

The dataset utilized, comprising speech, silence, and ambient sounds, provided a foundational testing ground. Yet, it may not fully represent the vast diversity of real-world acoustic scenarios. The system's effectiveness across different languages, dialects, and accents remains to be thoroughly evaluated.

\section{Conclusion}
In this paper, we have presented the development and implementation of an Voice Activity Detection (VAD) system engineered to excel in the presence of various ambient noises. Our system integrates innovative methodologies, including a specially designed filtering technique and optimized parameters, to accurately isolate speech amidst diverse background sounds. Through comprehensive testing and validation, we have demonstrated the robustness and precision of our VAD system in discerning speech from noise.

Our VAD system's success is underpinned by several key components and parameter choices. The 4th-order Butterworth filter, tailored to the frequency range of human speech (300 Hz to 1500 Hz), effectively removes unwanted background noise while preserving essential speech components. This filter's performance is further enhanced by carefully chosen parameters, including a window length of 0.31 seconds and a signal-to-noise ratio threshold of 90 dB. These parameters enable our system to accurately identify speech intervals amidst challenging acoustic environments, striking a balance between speech detection and noise suppression.

The experimental results, including the analysis of spectrograms, reinforce the effectiveness of our VAD system. The visual representation of speech, ambient noise, mixed signals, and detected speech signals in Figure 4 demonstrates the system's capability to differentiate speech from noise accurately. This visual evidence aligns with metrics, affirming the system's potential for real-world applications.

In conclusion, our VAD system, with its novel algorithm, filtering technique, and optimized parameters, represents a significant step towards achieving high accuracy in speech detection in noisy environments.
\section{Future Work}
As VAD technology becomes more prevalent in our daily lives, ethical considerations, such as user consent, data security, and algorithmic bias, must be addressed. Future research should explore ethical frameworks and guidelines for the responsible development and deployment of VAD systems.

Future research can focus on more advanced noise modeling techniques that can adapt dynamically to changing acoustic environments. Machine learning approaches, such as those discussed in [4], [5], and [6] can be explored to develop noise models that can better suppress complex and non-stationary background noises.
\section*{Broader Impact}
As we move forward, the broader implications of VAD technology, including privacy, security, and user autonomy, must be carefully considered. Striking a balance between enhanced functionality and respect for user rights remains a critical challenge. Future work may delve into addressing these considerations while further refining VAD systems to meet the evolving needs of a connected world.
\section*{Acknowledgment}
I would like to thank Dr. John Carmody for his helpful instruction on the topic of VAD systems and his expertise in the field of audio signal processing.

\section*{Author Information}
\textbf{Joshua Ball}, Graduate Student, Department of Electrical and Computer Engineering, Johns Hopkins University 

\vspace{12pt}

\end{document}